\documentclass[english,latin9,preprint, authoryear, trackchanges]{aa}
\usepackage[T1]{fontenc}
\usepackage{verbatim}
\usepackage{float}
\usepackage{textcomp}
\usepackage{amsmath}
\usepackage{graphicx}
\usepackage{subscript}

\makeatletter
\usepackage{lineno}
\usepackage{graphicx}
\usepackage{txfonts}
\usepackage{siunitx}
\usepackage{hyperref}

\newcommand*\patchAmsMathEnvironmentForLineno[1]{
  \expandafter\let\csname old#1\expandafter\endcsname\csname #1\endcsname
  \expandafter\let\csname oldend#1\expandafter\endcsname\csname end#1\endcsname
  \renewenvironment{#1}
  {\linenomath\csname old#1\endcsname}
  {\csname oldend#1\endcsname\endlinenomath}}
  \newcommand*\patchBothAmsMathEnvironmentsForLineno[1]{
  \patchAmsMathEnvironmentForLineno{#1}
  \patchAmsMathEnvironmentForLineno{#1*}}
  \AtBeginDocument{
  \patchBothAmsMathEnvironmentsForLineno{equation}
  \patchBothAmsMathEnvironmentsForLineno{align}
  \patchBothAmsMathEnvironmentsForLineno{flalign}
  \patchBothAmsMathEnvironmentsForLineno{alignat}
  \patchBothAmsMathEnvironmentsForLineno{gather}
  \patchBothAmsMathEnvironmentsForLineno{multline}
}

\makeatother

\usepackage{babel}

\begin{document}

\title{No detection of SO\textsubscript{2}, H\textsubscript{2}S, or OCS
in the atmosphere of Mars from the first two Martian years of observations
from TGO/ACS}\titlerunning{}
\author{A.~S.~Braude\inst{1}\and F.~Montmessin\inst{1}\and K.~S.~Olsen\inst{2}\and A.~Trokhimovskiy\inst{3}\and O.~I.~Korablev\inst{3}\and F.~Lef\`evre\inst{1}\and A.~A.~Fedorova\inst{3}\and J.~Alday\inst{2}\and L.~Baggio\inst{1}\and A.~Irbah\inst{1}\and G.~Lacombe\inst{1}\and F.~Forget\inst{4}\and E.~Millour\inst{4}\and C.~F.~Wilson\inst{2}\and A.~Patrakeev\inst{3}\and A.~Shakun\inst{3}}
\institute{Laboratoire Atmosph\`eres, Milieux, Observations Spatiales (LATMOS),
UVSQ Universit\'e Paris-Saclay, Sorbonne Universit\'e, CNRS, Paris,
France \\
\email{ashwin.braude@latmos.ipsl.fr}\and AOPP, Oxford University,
Oxford, United Kingdom\and Space Research Institute (IKI) RAS, Moscow,
Russia\and Laboratoire de M\'et\'eorologie Dynamique/IPSL, Sorbonne
Universit\'e, ENS, PSL Research University, Ecole Polytechnique,
CNRS, Paris, France}
\date{Received ...; Accepted...}
\titlerunning{No sulphur compounds in the Martian atmosphere}
\abstract{The detection of sulphur species in the Martian atmosphere would be
a strong indicator of volcanic outgassing from the surface of Mars.}{We wish to establish the presence of SO\textsubscript{2}, H\textsubscript{2}S,
or OCS in the Martian atmosphere or determine upper limits on their
concentration in the absence of a detection.}{We perform a comprehensive analysis of solar occultation data from
the mid-infrared channel of the Atmospheric Chemistry Suite instrument,
on board the ExoMars Trace Gas Orbiter, obtained during Martian years
34 and 35.}{For the most optimal sensitivity conditions, we determine 1$\sigma$
upper limits of SO\textsubscript{2} at 20 ppbv, H\textsubscript{2}S
at 15 ppbv, and OCS at 0.4~ppbv; the last value is lower than any previous
upper limits imposed on OCS in the literature. We find no evidence
of any of these species above a 3$\sigma$ confidence threshold. We
therefore infer that passive volcanic outgassing of SO\textsubscript{2}
must be below 2~ktons/day.}{}

\maketitle

\section{Introduction}

Multiple pieces of evidence point to the presence of substantial past
volcanic activity on Mars.\ This activity may in fact have been key in maintaining a
sufficiently warm, dense, and moist palaeoclimate through greenhouse
gas emission, thus allowing for a stable presence of surface liquid water
(e.g. \citealp{craddockgreeley2009}). While this past activity is
partly reflected in gaseous isotope ratios that are found in the present-day
Martian atmosphere \citep{jakoskyphillips2001,craddockgreeley2009},
more obvious evidence is provided by the numerous large volcanic structures
and lava flows present over much of the surface of Mars that have been
studied since the first spacecraft observations made by the Mariner
9 orbiter in the 1970s \citep{masursky1973}. Evidence of present-day
outgassing activity on Mars, however, remains more elusive. While
no signs of thermal hotspots were found by the Mars Odyssey/Thermal Emission Imaging System (THEMIS) instrument \citep{christensen2003themis},
evidence from the Mars Express/High Resolution Stereo Camera (HRSC)
showed signs of very intermittent activity from volcanoes that erupt
between periods of dormancy that last for hundreds of millions
of years around the Tharsis and Elysium regions \citep{neukum2004},
with the most recent activity in some cases dating from only a few
million years ago. More recently, the InSight Lander found that seismic activity was generally lower than expected on Mars; however, the two
largest marsquakes it managed to detect were in the Cerberus Fossae
region \citep{giardini2020insight}.\ This was later accompanied by evidence
from the Mars Reconnaissance Orbiter/High Resolution Imaging Experiment (HIRISE) of extremely recent volcanic activity
in the same region dating back only around 100,000 years \citep{horvath2021}.This
gives further credence to the possibility of residual volcanic activity
continuing to the present day, perhaps from localised thermal vents.

On Earth, sulphur dioxide (SO\textsubscript{2}) tends to be by far
the most abundant compound emitted through passive volcanic degassing
after CO\textsubscript{2} and water vapour; it is detected together with smaller
amounts of hydrogen sulphide (H\textsubscript{2}S) and occasionally
carbonyl sulphide (OCS), although the exact ratios
of these compounds usually depend on the volcano and the type of
eruption or outgassing in question \citep{Symonds1994}. \citet{gaillard2009} used constraints
on the composition of Martian basalts, together with thermochemical
constraints on the composition of the Martian mantle, to estimate
a sulphur content of potential Martian volcanic emission, especially
from volcanoes that formed later in Mars' history, such as those in
the Tharsis range, which is 10 - 100 times higher than equivalent sulphur
emission from Earth volcanoes. Since these sulphurous gases have stabilities
in the Martian atmosphere of the order of only days to a few years
\citep{nair1994,krasnopolsky1995,wong2003,wong2005}, their detection
in the Martian atmosphere would be a strong indicator of residual
present-day volcanic activity on Mars. Alternatively, it could be
a sign of a coupling between the atmosphere and sulphur-containing
regoliths \citep{farquhar2000} or sulphate deposits that are
present in multiple regions of Mars \citep{bibring2005,langevin2005}. This would be
an indirect sign of past volcanic activity but one that would nonetheless
shed light on a previously unknown sulphur cycle in the Martian atmosphere.

Recent detections of other new trace gases in the atmosphere of Mars
from remote sensing provide a tantalising glimpse of fascinating new
Martian surface and atmospheric chemistry that until now had remained
unknown to science. Examples include the independent detections of
hydrogen peroxide by \citet{clancy2004htwootwo} and \citet{encrenaz2004},
a photochemical product that acts as a possible sink for organic compounds
in the atmosphere; and the independent detections of hydrogen chloride
(HCl) reported by \citet{korablev2020hcl}, \citet{olsen2021hcl},
and \citet{aoki2021}, a possible tracer of the interaction of dust
with the atmosphere. HCl in particular was historically assumed to
be a tracer of volcanic outgassing and was hypothesised to be responsible
for the observed geographical distribution of elemental chlorine \citep{keller2006}
and perchlorates \citep{hecht2009,catling2010}. While the seasonality
of observed HCl detections in the Martian atmosphere has made a volcanic
origin less likely, more recent anomalous detections of HCl in aphelion
reported by \citet{olsen2021hcl} have not entirely ruled it out.
We should of course also note various disputed detections of methane
(e.g. \citealp{formisano2004methane,krasnopolsky2004,mumma2009,webster2015methane}),
whose ultimate provenance remains subject to debate and could be a
result of a number of different processes, including volcanic outgassing
(\citealt{oehleretiope2017}; and references therein), but whose legitimacy
has been seriously questioned by recent satellite measurements \citep{korablev2019,montmessin2021}.%
\begin{comment}
Observations of sulphurous gases concurrently with future methane
detections could therefore constrain whether the methane has a biological
or an abiological origin. In spite of this
\end{comment}

Despite this, all attempts to constrain the presence of sulphurous
gases in the Martian atmosphere, and in particular SO\textsubscript{2},
have failed to confirm any statistically significant positive detections.
Although comprehensive searches for trace gases in the Martian atmosphere
have been ongoing since the Mariner 9 era \citep{maguire1977}, most
of the earliest attempts were hampered by the lack of spectral sensitivity
and low spectral resolution: the resolving power in the mid-infrared
where many of the required spectral signatures are located generally
needs to be greater than 10,000 in order to resolve them from neighbouring
CO\textsubscript{2} and H\textsubscript{2}O absorption lines. More
recent literature concerning upper limits relied on ground-based observations
that were made at much higher spectral resolutions, with SO\textsubscript{2}
and H\textsubscript{2}S spectral data obtained in the sub-millimetre (e.g.
\citealp{nakagawa2009,encrenaz2011so2}) and the thermal infrared
\citep{krasnopolsky2005,krasnopolsky2012upperlims} wavenumber ranges,
and corresponding OCS spectral data around an absorption band in the
mid-infrared \citep{khayat2017}. However, ground-based observations
still have a number of drawbacks. Firstly, they are hindered by a
lack of temporal coverage. Secondly, the presence of telluric absorption
in ground-based observations also obscures gas absorption signatures
of Martian provenance, removal of which requires correction for Doppler
shift (cited by \citealt{zahnle2011} as a possible source of false
detections of methane) as well as a complex radiative transfer model
that decouples the contribution from Earth and Mars. Finally, they
can often lack the spatial resolution necessary to resolve local sources
of emission. By contrast, observations from probes in orbit suffer
less from these issues and have more localised spatial coverage, which
enabled the recent detection of HCl in orbit despite numerous recent
failed attempts to detect it from the ground, even with instruments
of similar spectral resolution (e.g. \citealp{krasnopolsky1997,hartogh2010,villanueva2013}).

In this work we present upper detection limits on SO\textsubscript{2},
H\textsubscript{2}S, and OCS based on spectral data from the first two Martian years of observations from the mid-infrared channel of the Atmospheric Chemistry Suite instrument \citep[ACS MIR;][]{korablev2018} on board the ExoMars Trace Gas Orbiter (TGO; \citealp{Vago2015}). In Sect. 2 we introduce the
ACS MIR data and summarise the data calibration process, and in
Sect. 3 we present the two methods used to establish detection limits
on each of the given compounds. In Sect. 4 we present the retrieved
upper limits for each of the sulphur trace species in turn. Discussion
and conclusions are reserved for Sect. 5.

\section{Data and calibration}

TGO has been in orbit around Mars since October 2016 and has operated
continuously since the start of its nominal science phase in April
2018, providing a wealth of data covering one and a half Martian years
as of writing (from L\textsubscript{s}= 163$\text{\textdegree}$ in MY 34 to
L\textsubscript{s}= 352$\text{\textdegree}$ in MY 35). On board are two sets
of infrared spectrometers designed to perform limb, nadir, and solar
occultation observations of the atmosphere: the ACS (\citealp{korablev2018}) and the Nadir and Occultation
for Mars Discovery (NOMAD; \citealp{vandaele2018}) instruments. In
this work, we focus on the ACS MIR instrument (\citealp{trokhimovskiy2015mir}),
which observes Mars purely through solar occultation viewing geometry
with the line of sight parallel to the surface, obtaining a set of
transmission spectra of the Martian atmosphere at individual tangent
heights separated at approximately 2 km intervals from well below
the surface of Mars up to the top of the atmosphere during a single
measurement sequence. The closer the tangent height of observation
to the surface, the greater the molecular number density of the absorbing
species and hence the greater the optical depth integrated along the
line of sight. This allows for higher sensitivities to very small
abundances of trace gases, in theory down to single parts per trillion
for some species \citep{korablev2018,toon2019}, than can be achieved
from ground-based observations where the path length through the atmosphere
is much shorter. In practice, however, the sensitivity is limited
by instrumental noise and artefacts, and the signal-to-noise ratio
(S/N) usually decreases at lower altitudes due to the attenuation
of the solar signal by the presence of dust, cloud and gas absorption.
The best upper limits for trace gas species are therefore usually
obtained as close to the surface as possible under very clear atmospheric
conditions, which occur especially near the winter poles, during aphelion
and excluding global dust storm periods.

\begin{figure}[h]
\includegraphics[bb=20bp 0bp 578bp 248bp,width=1.1\columnwidth]{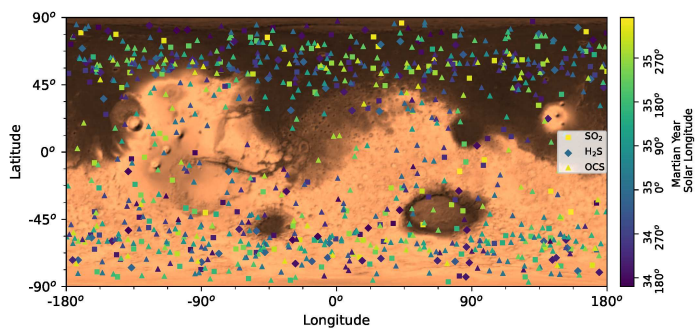}

\caption{Spatial and temporal distribution of all the ACS MIR observations
covered in this analysis. Squares represent observations made by ACS
using grating position 9 (sensitive to SO\protect\textsubscript{2}),
diamonds using grating position 5 (H\protect\textsubscript{2}S), and
triangles using grating position 11 (OCS). The colour of each symbol
represents the time, in units of solar longitude ($L_{s}$) over Martian
years (MY) 34 and 35, at which the observation was obtained. The relative
concentration of observations made closer to the poles is due to the
TGO orbital geometry. Latitude values are planetocentric. The background
image of Mars is based on topography data from the Mars Orbiter Laser Altimeter (MOLA) instrument
on board Mars Global Surveyor \citep{smith2001mola}.}

\label{spatialdistribution}
\end{figure}

ACS MIR is a cross-dispersion echelle spectrometer that covers a spectral
range of 2400-4500~cm\textsuperscript{-1} with a spectral resolving
power of up to 30,000 \citep{trokhimovskiy2020sashaslines}. The instrument
makes use of a steerable secondary reflecting grating that then disperses
the incoming radiation and separates it into named diffraction orders
that encompass smaller wavenumber subdivisions. The group of diffraction
orders that is selected depends on the position of the secondary grating.
In this work we analyse observations obtained using three grating
positions: position 9, which covers the main SO\textsubscript{2} absorption
bands; position 11, which covers the strongest OCS absorption bands; and
position 5, which covers H\textsubscript{2}S. Respectively, these
grating positions cover wavenumber ranges of 2380 - 2560~cm\textsuperscript{-1}
(diffraction orders 142 - 152), 2680 - 2950~cm\textsuperscript{-1}
(160 - 175), and 3780 - 4000~cm\textsuperscript{-1} (226 - 237).
In each case, only one grating position is used per measurement sequence
and the spatial distribution of these measurements over the time period
covered in this analysis is shown in Fig. \ref{spatialdistribution}.
The end product is a 2D spectral image from each tangent height projected
onto a detector, which consists of wavenumbers dispersed in the x axis 
and diffraction orders separated along the y axis (the reader is directed
to \citet{korablev2018,trokhimovskiy2020sashaslines,olsen2020co,montmessin2021}
for diagrams and further discussion). This image contains approximately
20 - 40 pixel rows per diffraction order depending on the grating
position in question, corresponding to an increment of around 200
- 300 m in tangent height per row. The intensity distribution of incoming
radiation is distributed asymmetrically over each diffraction order,
with the row of maximum intensity, and hence maximum S/N, usually
located somewhere between the geometric centre of the diffraction
order and the edge of the slit that is over the solar disc.

We should also note the presence of `doubling', an artefact of unknown
origin that causes two images per tangent height, offset horizontally
and vertically by a few pixels, to be projected onto the detector
surface, resulting in transmission spectra in which each absorption
line appears to be divided into two separate local minima \citep{alday2019,olsen2020co}.
This doubling effect tends to be strongest in the centre of the illuminated
portion of the slit where the signal strength is highest, and its
profile changes across each row of a given diffraction order in ways
that are still under investigation as of writing. In practice it also
acts to reduce the effective spectral resolution of ACS MIR spectra
from the nominal value \citep{olsen2021hcl} and hence proves a major
source of error in the determination of upper limit values. 

Calibration was performed according to the procedure detailed in \citet{trokhimovskiy2020sashaslines}
and \citet{olsen2020co}. Observations in the measurement sequence
when the field-of-view is fully obscured by Mars were used to estimate
dark current and thermal background. At the other end of the measurement
sequence, observations in the very high atmosphere were used to estimate
the solar spectrum, taking account the drift of the image on the detector
over time induced by the thermal background. Stray light was estimated
from the signal level between adjacent diffraction orders. A first
guess of the pixel-to-wavenumber registration was made by comparing
the measured solar spectrum with reference solar lines \citep{hase2010},
which was then further refined as part of the retrieval process as
will be described in Sect. \ref{subsec:RISOTTO}. The tangent height
of each observation at the edges of the slit was estimated using the
TGO SPICE kernels \citep{trokhimovskiy2020sashaslines}, where the
row positions of the top and bottom of each slit were established
using reference `sun-crossing' measurement sequences in which the
slit passes across the Sun perpendicularly to the incident solar radiation
path above the top of the atmosphere. Nonetheless, there is still some
uncertainty on the tangent height registration as a) although we assume
a linear distribution, the change in tangent height over each intermediate
slit row is not well constrained and b) the pixel offset between the
two doubled images can make it difficult to identify the exact locations
of the slit edges on the detector array. This uncertainty is usually
of the order of around 0.5 - 1.0~km. %
\begin{comment}
These do not currently exist for grating position 11, and so cruder
estimates of the locations of the slit edges must be made based on
the shape of the intensity profile across the slit.
\end{comment}

\section{Analysis\label{sec:Method}}

\subsection{RISOTTO forward model\label{subsec:RISOTTO}}

\begin{figure*}
\includegraphics[width=1\textwidth]{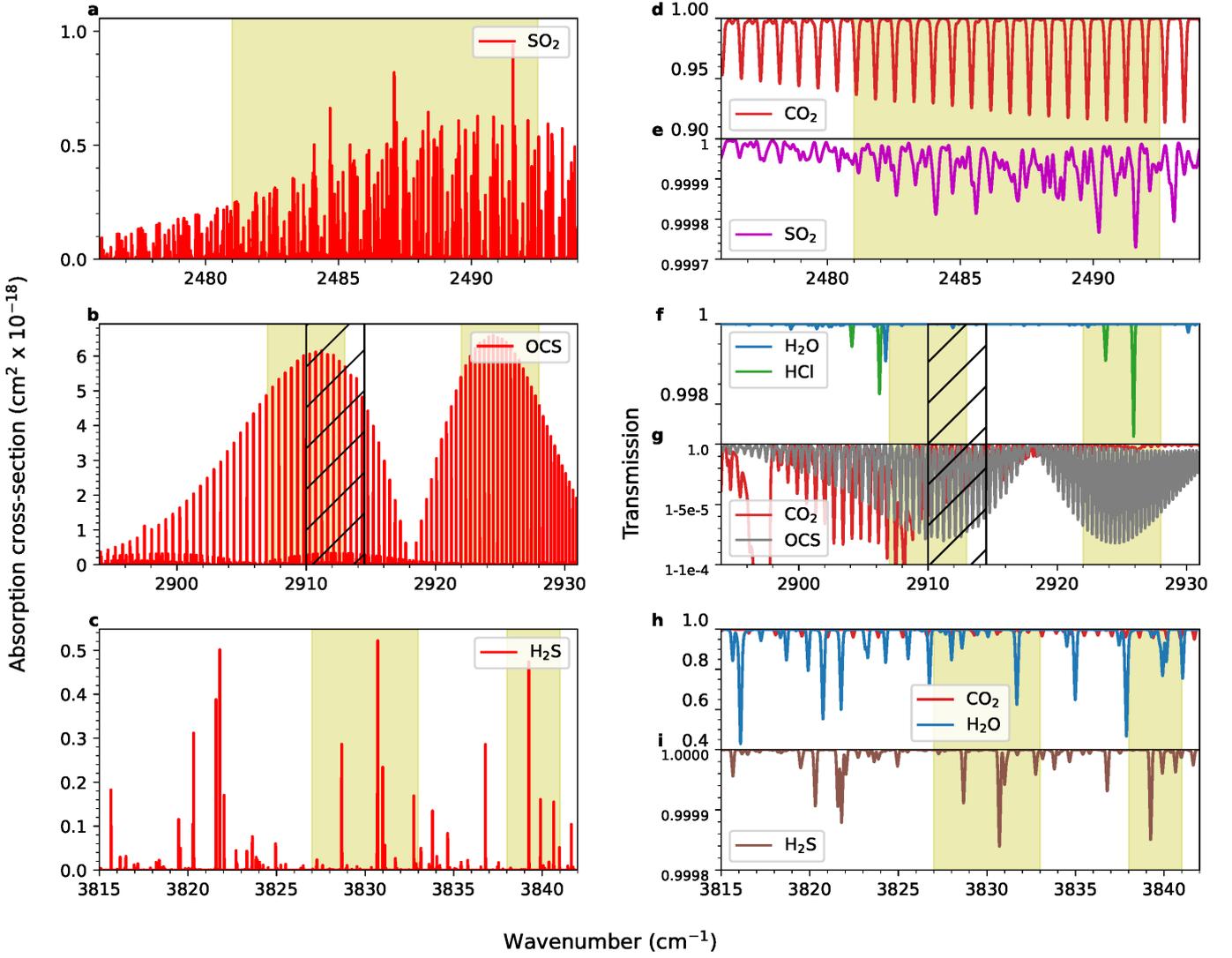}

\caption{Computed cross-section values and simulated transmission spectra,
for illustrative reference pressure and temperature values of 1 mbar
and 200 K, respectively. Displayed wavenumber ranges reflect the coverage
of the analysed diffraction orders, with the light yellow highlighted
regions showing the spectral intervals over which detection limit retrievals
were conducted, maximising absorption from the trace species in question
while minimising overlapping absorptions from other known species.
\emph{Panels a-c: }Cross-section values for \emph{(a) }SO\protect\textsubscript{2}
in diffraction order 148, \emph{(b)} OCS in diffraction orders 173
and 174, and \emph{(c)} H\protect\textsubscript{2}S in diffraction
order 228. The grey cross-hatched regions in panels \emph{(b), (f),
}and \emph{(g) }show the wavenumber range overlap between diffraction
order 173 on the left and diffraction order 174 on the right. \emph{Panels
d-i:} Estimated transmission spectra given the aforementioned pressure
and temperature conditions, convolved with a Gaussian instrument function
of resolution R $\sim$ 20,000. For each diffraction order, we plot
the contributions of the trace species to the total transmission separately
from the other interfering species, for which the contribution is several
orders of magnitude larger.}
\label{synthetic}
\end{figure*}

Upper limits are obtained using the RISOTTO radiative transfer and
retrieval pipeline \citep{braude2021soar}. RISOTTO relies on Bayesian
optimal estimation, starting from a prior state vector of parameters
relating both to the atmospheric quantities that we wish to retrieve
from the data (in this case, vertical gas abundance profiles quantified
in units of volume mixing ratio, hereon abbreviated to VMR) and a
number of instrumental parameters intended to reduce systematic errors
due to noise, uncertainties in instrumental calibration or the presence
of aerosol. We model three of these instrumental parameters specifically:
first, a polynomial law that relates the first guess of the pixel-to-wavenumber
registration to a more accurate registration; second, an instrument
line shape model akin to that described by \citet{alday2019} that
approximates the doubling artefact observed in spectra from ACS MIR;
and third, the transmission baseline as a function of wavenumber that
takes into account broad variations in the shape of the spectrum while
simultaneously avoiding problems of so-called overfitting, where
any narrow residuals due to individual noise features or poorly modelled
gas absorption lines are compensated for by changing the baseline
in an unphysical manner. The algorithm then computes a forward model
based on these parameters and then iteratively finds the optimal values
of both the scientific quantities and the instrumental parameters
simultaneously that best fit the observed transmission spectra, together
with their associated uncertainties.%
\begin{comment}
, which acts to retrieve a vertical profile of a given gaseous species
in the atmosphere given a measurement sequence and a single row per
wavenumber window, for which the retrieval of multiple separate windows
per single diffraction order are permitted, as well as the simultaneous
retrieval of multiple windows over separate diffraction orders. The
retrieval also acts to iteratively correct for three major instrumental
uncertainties: the variable transmission baseline due to the presence
of noise and aerosol absorption, the instrument line shape function
associated with the aforementioned doubling which is approximated
using two Gaussians of finite resolution similar to the parametrisation
in \citet{alday2019}, and small uncertainties in the spectral registration.
\end{comment}
{} In order to model gas absorption along the line of sight both accurately
and in a manner that is computationally efficient, spectral absorption
cross-section lookup tables are calculated from the HITRAN 2016 database
\citep{GORDON2017} over a number of sample pressure and temperature
values that reflect the range of values typically found in the Martian
atmosphere, and then a quartic function is derived through regression
to compute the cross-sections at any given intermediate pressure and
temperature value as explained further in \citet{braude2021soar}.
Examples of these computed cross-sections for each of the three molecules
studied in this article are shown in Fig. \ref{synthetic}a-c, with
the corresponding wavenumber selections for the retrievals set to
favour stronger molecular lines while avoiding lines that overlap
too greatly with lines of other known molecules present in the region.
As an illustration of how these species may appear in real ACS MIR
spectra given predicted abundances, Fig. \ref{synthetic}d-i show
equivalent approximate simulated transmissions for the major species
in the spectral ranges present in the three orders, with conditions
equivalent to those near the south pole at aphelion around 10 km of
altitude, and convolved with a Gaussian instrument line shape function
of R~$\sim$~20,000 to roughly reflect the reduction in spectral
resolution due to doubling. For both figures we assume an atmosphere
of 95.5\% CO\textsubscript{2}, 100~ppmv H\textsubscript{2}O, 10~ppbv
of SO\textsubscript{2} and H\textsubscript{2}S, and 1~ppbv of HCl
and OCS, corresponding either to simulated detection limits in optimal
dust conditions \citep{korablev2018} or existing measurements of
known species (eg. \citealt{fedorova2020,korablev2020hcl}). We can
see that the regions in which the strongest SO\textsubscript{2} and
H\textsubscript{2}S lines are located are dominated by much stronger
lines of CO\textsubscript{2} and H\textsubscript{2}O. By contrast,
the spectral region in which OCS is found is usually much clearer,
with HCl absorption lines easily resolvable and only found during
certain seasons.

At the wavenumber ranges studied in this analysis, the CO\textsubscript{2}
absorption lines are not usually strong enough to independently constrain
gaseous abundances, temperature and pressure all simultaneously. This
can be even further exacerbated by uncertainties induced by systematic
errors in the instrument line shape due to doubling (e.g. \citealp{alday2021co2}).
Nonetheless, retrieved gas abundances are heavily degenerate with
temperature and pressure and so completely arbitrary values cannot
be chosen. \emph{A priori }temperature and pressure data are therefore
usually sourced from fitting stronger CO\textsubscript{2} bands in
the near-infrared (NIR) channel of ACS data obtained during the same
measurement sequence \citep{fedorova2020}. For occasional ACS measurement
sequences, however, these observations are not available, in which
case we use estimates from the Laboratoire de M\'et\'eorologie Dynamique general circulation
model (LMD GCM) solar occultation database
\citep{forget1999mcd,millour2018mcd,forget2021lmdgcm}, which computes
vertical profiles of a number of atmospheric parameters for the given
time and location of an ACS measurement sequence using a GCM. For retrievals where no CO\textsubscript{2} bands are
to be fitted, such as with OCS, it is usually adequate to leave the
temperature-pressure profile fixed in the retrieval. For retrievals
of SO\textsubscript{2} and H\textsubscript{2}S where they are to
be fitted, the LMD GCM estimates are not always close enough to the
true value to provide a good fit to the CO\textsubscript{2} absorption
bands, and so minor adjustments in the temperature-pressure profile
do have to be made in the retrieval. To decouple the effects of temperature
and pressure, we keep a reference pressure value at a given altitude
fixed to that from the LMD solar occultation database, then use the
CO\textsubscript{2} bands to retrieve a vertical temperature profile
directly from the data and thereby derive a vertical pressure profile
by assuming hydrostatic equilibrium (e.g. \citealp{quemerais2006,alday2019,alday2021fractionation,montmessin2021}).

\subsection{Retrieval procedure}

\begin{comment}
When we attempted to apply this method to retrievals of SO\textsubscript{2}
however, we found that the degeneracy with the baseline level due
to high continuum absorption of SO\textsubscript{2}, together with
the presence of the most prominent SO\textsubscript{2} signatures
near the edge of the detector where the SNR and uncertainties in calibration
are highest, resulted in the retrieval of unrealistically low upper
limits as well as a large number of false detections. This therefore
forced us to use a different, more empirical method for SO\textsubscript{2}.
We discuss these two methods in turn in the next two subsections.
\end{comment}
{} %
\begin{comment}
figure with retrieval of profile given different injections of the
trace species abundance
\end{comment}

\subsubsection{H\protect\textsubscript{2}S and OCS\label{subsec:HS-and-OCS}}

A number of approaches can be used to derive the upper limit of detection
of a given species, each with their advantages and disadvantages.
The most common approach is to define the upper limit as the estimate
of the uncertainty on a retrieved VMR given a true VMR equal to 0;
if the retrieved VMR was greater than this uncertainty value, it would
be deemed a detection as opposed to an upper limit. This uncertainty
can be estimated either by quantifying each of the sources of error
in turn (e.g. \citealp{aoki2018}) or by deriving a first estimate
of the\emph{ }uncertainty from a retrieval code and then adjusting
the uncertainties \emph{post hoc }according to statistical criteria
(e.g. \citealp{montmessin2021,knutsen2021}). This method is expanded
on by \citet{olsen2021ph3} by performing an independent retrieval
of the vertical gas profiles using multiple pixel rows on the ACS
MIR detector array, and then calculating an upper limit value from
the standard error on the weighted average of the vertical profiles
from each of these retrievals combined. This has the added benefit
of increasing the statistical significance of a measurement through
repeated observation and thereby deriving lower upper limits%
\begin{comment}
, however we find in our own retrievals that it is also sensitive
to intrinsic biases imposed either by the prior state vector or by
local artefacts in the spectrum
\end{comment}
.

\begin{figure}[h]
\includegraphics[width=1\columnwidth]{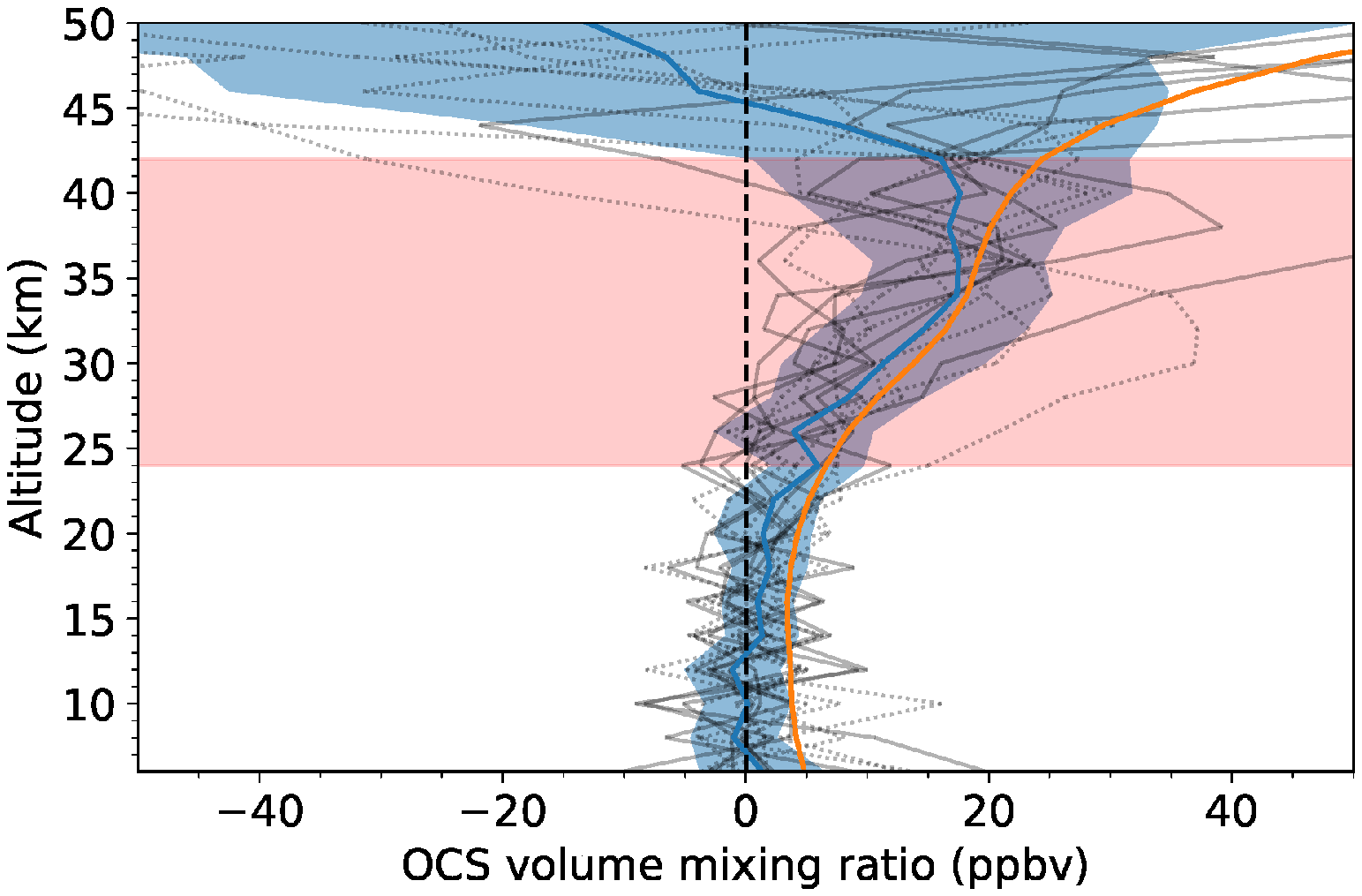}

\caption{Illustrative example of the upper limit derivation methodology for
OCS and H\protect\textsubscript{2}S. Grey lines indicate the individual
$2N$ vertical profiles of OCS retrieved according to \emph{(solid)
}step 3 and \emph{(dotted)} step 4, for each row on the detector array.
The solid blue line shows the weighted mean abundance profile, $\mu_{j}^{*}$,
derived from the individual retrievals, with the standard deviation,
$\varsigma_{j}^{*}$, shown in the shaded blue region. The solid orange
line shows the profile of $\sqrt{\mu_{j}^{*}+\varsigma_{j}^{*}}$
smoothed over altitude. The red shaded region shows the altitude range
for which a 1$\sigma$ detection was determined, i.e. where the value
of $\varsigma_{j}^{*}$ is lower than the value of $\mu_{j}^{*}$.}

\label{methodology_visual_example}
\end{figure}

For retrievals of upper limits of OCS and H\textsubscript{2}S, we
use a modified version of the \citet{olsen2021ph3} method, but with
some further amendments to minimise uncertainties in the estimated
spectral noise level, as well as to reduce sensitivity due to intrinsic
biases imposed either by the prior state vector or by local artefacts
in the spectrum. The sigma detection value at each altitude is determined
by the ratio of the weighted mean abundance value $\mu_{j}^{*}$ to
the weighted standard deviation $\varsigma_{j}^{*}$ as found by the
procedure detailed in Appendix \ref{sec:Detailed-procedure-of}, that
is to say, where $\mu_{j}^{*}/\varsigma_{j}^{*}=1$ indicates a positive
1$\sigma$ detection. Usually, a 1$\sigma$ or 2$\sigma$ detection
indicates overfitting of local artefacts or noise features as opposed
to a genuine positive detection. We therefore treat anything below
a 3$\sigma$ detection as insignificant and define our upper limit
at each altitude by adding total sources of both systematic and random
error in quadrature, that is to say, the upper limit is equal
to $\sqrt{\mu_{j}^{*2}+\varsigma_{j}^{*2}}$ smoothed using a narrow
Gaussian filter with respect to altitude, as shown in Fig. \ref{methodology_visual_example}.

\subsubsection{SO\protect\textsubscript{2}\label{subsec:SO}}

Retrievals of SO\textsubscript{2} are complicated by the fact that
SO\textsubscript{2} has a dense line structure that is difficult
to fully resolve at the given spectral resolution, and so retrievals
of SO\textsubscript{2} are somewhat difficult to decouple from uncertainties
in the baseline level induced by noise, calibration errors, and aerosol
extinction. Failure to take into account uncertainties in the baseline
level will therefore result in upper limits of SO\textsubscript{2}
that are too low. While RISOTTO should ordinarily be able to handle
some degeneracy with low frequency baseline variations \citep{braude2021soar}
that would save having to perform a more complicated fitting procedure
as in \citet{belyaev2008,belyaev2012} to fully separate out the contribution
of aerosol and SO\textsubscript{2} to the continuum absorption, the
situation is made complicated by the presence of artefacts near the
edge of the detector where the degeneracy between low-frequency baseline
variations and SO\textsubscript{2} absorption can most easily be
broken due to the presence of prominent SO\textsubscript{2} lines.
The retrieval code therefore has the propensity of overfitting artefacts,
resulting in false detections of highly negative abundances of SO\textsubscript{2}.

One method is to start with a forward model where the species is absent,
and gradually inject incrementally large quantities of the species
into the forward model until the change in the spectral fit is deemed
to increase above a given threshold (eg. \citealp{teanby2009,korablev2020hcl}).
In practice this can become slow and unwieldy in the case of solar
occultations where there are multiple spectra to be fit simultaneously,
each with their own independent noise profile and sensitivity to the
given species. In addition, it also begs the question as to how the
threshold should be defined quantitatively, especially if the S/N
is uncertain. 

\citet{teanby2009} assessed the significance of a detection according
to the mean squared difference between the observed spectrum $y_{obs}(\nu)$
as a function of wavenumber, $\nu$, and the modelled spectrum $y_{mod}(\nu,x)$
given an `injected' gas volume mixing ratio, $x$, weighted according
to the spectral uncertainty $\sigma(\nu)$:

\begin{alignat}{1}
\Delta\chi^{2} & =\chi^{2}(x)-\chi^{2}(0)\\
 & =\sum_{i}^{N_{\nu}}\left(\left(\frac{y_{obs}(\nu_{i})-y_{mod}(\nu_{i},x)}{\sigma(\nu_{i})}\right)^{2}-\left(\frac{y_{obs}(\nu_{i})-y_{mod}(\nu_{i},0)}{\sigma(\nu_{i})}\right)^{2}\right),
\end{alignat}

 \noindent where $N_{\nu}$ is the number of individual spectral points in a
given observation. The residual errors on the fit to the spectrum
are assumed to follow a double exponential distribution \citep{press1992},
neglecting systematic errors:

\begin{equation}
P(y_{obs}(\nu_{i})-y_{mod}(\nu_{i},0))\sim\exp\left(\;\;\left|\frac{y_{obs}(\nu_{i})-y_{mod}(\nu_{i},0)}{\sigma(\nu_{i})}\right|\;\;\right)\label{eq:proberror}
.\end{equation}

\citet{teanby2009,teanby2019} then defined upper limits according
to the confidence interval around the mean of the probability distribution
described by Eq. \ref{eq:proberror}, so that values of gas abundance
$x$ that give a value of $\Delta\chi^{2}=n^{2}$ should represent
an \emph{n}-sigma upper limit, while a gas abundance that results
in $\Delta\chi^{2}=-n^{2}$ should analogously represent a positive
\emph{n}-sigma detection. However, this method was used for single
isolated lines where the detection limit is more or less independent
of the size of the spectral window. This is not the case for SO\textsubscript{2},
which has a dense line structure that affects the shape of the baseline,
and where there is substantial uncertainty on the values of $\sigma(\nu_{i})$
induced by various sources of systematic error, notably from the doubled
instrument line shape. We find in our own data that values of $\Delta\chi^{2}=1$
give detection limits that are far too optimistic even when correcting
for spectral sampling as in \citet{teanby2019}. Instead, we find
that taking the standard deviation of the double exponential distribution
provides more realistic detection limits, which would therefore correspond
to \emph{n}-sigma upper limits and detections of, respectively, $\Delta\chi^{2}=\pm\sqrt{2N_{\nu}}n^{2}$.
Depending on the occultation in question, $\sqrt{2N_{\nu}}$ is usually
equal to around 25 for the wavenumber range taken into account in
the retrieval.

We first performed an initial retrieval where SO\textsubscript{2} is
not present in the forward model, only retrieving the temperature
profile and the three instrumental parameters as previously described,
to give a spectral fit $y_{mod}(\nu,0)$. The values of $\sigma(\nu)$
are then estimated by taking a moving average of the difference between
$y_{obs}(\nu)$ and $y_{mod}(\nu,0)$, thereby providing a first guess
of spectral uncertainties due to systematics in the spectra and forward
modelling error, and hence allowing a preliminary value of $\chi^{2}(0)$
to be calculated. Fixed vertical profiles of SO\textsubscript{2}
abundance $x$ were then added to the forward model, and a new retrieval
performed where the baseline is allowed to vary but the remaining
parameters of the state vector are fixed to those retrieved from $y_{mod}(\nu,0)$.
If this results in a value of $\Delta\chi^{2}$ at a given tangent
height that is negative, the values of $\sigma(\nu)$ are further
refined by taking a moving average of the difference between $y_{obs}(\nu)$
and $y_{mod}(\nu,x)$, so that the contribution of forward modelling
error to the spectral uncertainty is minimised. Once enough different
vertical profiles of SO\textsubscript{2} were modelled to sufficiently
sample the $\Delta\chi^{2}$ parameter space, the approximate 1$\sigma$
level at each altitude was found through progressive quadratic interpolation
to a volume mixing ratio profile where $\Delta\chi^{2}$ is as close
to $\sqrt{2N_{\nu}}$ as possible for each altitude, or conversely
where $\Delta\chi^{2}$ is as low as possible. Due to the presence
of systematic errors, we do not attempt to reduce upper limits by
adding the contribution of multiple rows of each diffraction order
as with H\textsubscript{2}S or OCS, instead only analysing a single
row close to the centre of the diffraction order that receives the
most input radiance and hence has the highest S/N.

\section{Results}

\subsection{SO\protect\textsubscript{2}}

Given that volcanic emissions of SO\textsubscript{2} are predicted
to dwarf those of any other sulphur species and have the longest photochemical
lifetime%
\begin{comment}
cite
\end{comment}
, SO\textsubscript{2} is the tracer that has attracted the most attention
in the literature out of all the three gases analysed here. Ground-based
observations of SO\textsubscript{2} in the past have focussed on
two main spectral regions. In the sub-millimetre  there is a single line present
at 346~GHz, from which disc-integrated upper limits of 2~ppbv were
derived by \citet{nakagawa2009} and upper limits of just over 1 ppbv
by \citet{khayat2015,khayat2017}. A number of rotational-vibrational
lines are also present in the thermal infrared between 1350 - 1375~cm\textsuperscript{-1}
\citep{encrenaz2004,encrenaz2011so2,krasnopolsky2005,krasnopolsky2012upperlims},
with upper limits of 0.3~ppbv independently confirmed by both \citet{encrenaz2011so2} and \citet{krasnopolsky2012upperlims}. By contrast, the strongest SO\textsubscript{2}
absorption band in the ACS MIR wavenumber range, centred around 2500~cm\textsuperscript{-1},
is still weaker than the absorption bands found in these two regions.
In addition, the SO\textsubscript{2} absorption band has a very dense
line structure, which at ACS MIR resolution results in only a small
number of isolated lines that are sufficiently prominent to decouple
the contributions of SO\textsubscript{2} absorption to the spectrum
from the uncertainty in the transmission baseline. This is also further
complicated by the fact that several lines overlap strongly with the
absorption bands of three separate isotopes of CO\textsubscript{2},
several of which are also difficult to resolve from one another and
contribute to the baseline uncertainty. The SO\textsubscript{2} line
that is best resolved from the baseline and CO\textsubscript{2} absorption
is present at 2491.5~cm\textsuperscript{-1}, but this is also located
close to the edge of the diffraction order where the S/N starts to
decrease. 

We performed retrievals on a relatively broad spectral range of 2481
- 2492~cm\textsuperscript{-1}, which allowed the strongest lines
of \textsuperscript{12}C\textsuperscript{16}O\textsuperscript{18}O
to be fit in order to constrain both the vertical temperature profile
and the doubling line shape, as well as minimising the probability
of local noise features at 2491.5~cm\textsuperscript{-1} being overfit
with SO\textsubscript{2} absorption, while avoiding lower wavenumbers
where the contribution of \textsuperscript{12}C\textsuperscript{16}O\textsuperscript{16}O
and \textsuperscript{12}C\textsuperscript{16}O\textsuperscript{17}O
to the uncertainty in the transmission baseline starts to become significant.
In addition, we also only fit spectra up to a tangent height of 50~km
above the Martian geoid (usually referred to as the `areoid')
as noise features start to dominate in spectra above this level, which
can occasionally be confused with genuine SO\textsubscript{2} absorption
by the retrieval code. This is justified as SO\textsubscript{2} emitted
from the surface would only be detectable at very high abundances
above 50~km as we show later in this section. 

\begin{figure}[h]
\includegraphics[width=1\columnwidth]{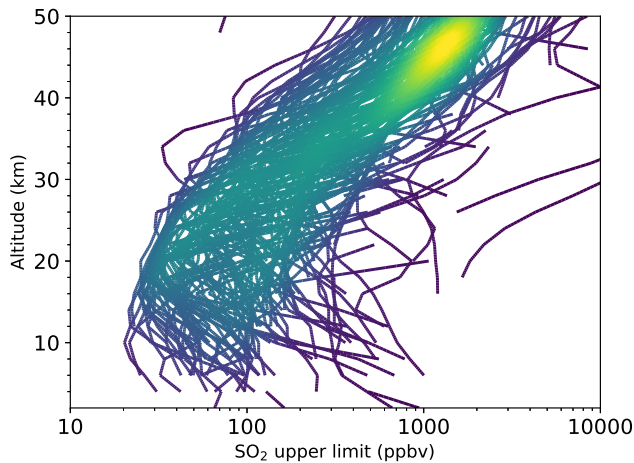}

\caption{1$\sigma$ upper limit values of SO\protect\textsubscript{2} obtained
as a function of altitude for each position 9 measurement sequence.
Yellower colours indicate greater densities of upper limit values
for a given altitude.}

\label{so2scatter}
\end{figure}

We retrieved estimates of upper limits of SO\textsubscript{2} from
all measurement sequences obtained using grating position 9 for which
the observed altitude of aerosol saturation was below~50 km, according
to the procedure previously outlined in Sect. \ref{subsec:SO}.
190 occultations in total satisfied this criterion, covering a time
period of L\textsubscript{s} = 165 - 280$\text{\textdegree}$ of MY 34 and
L\textsubscript{s} = 140 - 350$\text{\textdegree}$ of MY 35, in both cases
equivalent to the time around perihelion and approaching early aphelion.
Occasionally, vertical correlations between spectra, together with
the breakdown in the approximation of a quadratic relationship between
volume mixing ratio and $\Delta\chi^{2}$ due to baseline degeneracy,
can result in failure to converge to a $\Delta\chi^{2}\approx\sqrt{2N_{\nu}}$
solution for certain tangent heights. The altitude ranges at which
these poor $\Delta\chi^{2}$ solutions are found are therefore removed
from the upper limit profiles before they are smoothed. In Fig. \ref{so2scatter},
we plot the smoothed vertical upper limit profiles retrieved from
all 190 occultations as a function of altitude. We find in most cases
that we can derive upper limits down to around 30 - 40~ppbv, usually
around 20~km above the areoid, with the sensitivity decreasing exponentially
with altitude until only measurements of the order of 1~ppmv of SO\textsubscript{2}
are retrievable in the lower mesosphere. Below the 20~km level, the
presence of aerosols usually lowers the S/N to the point where the
CO\textsubscript{2} lines become heavily distorted, particularly
below around 10 - 15~km, which makes accurate fitting of SO\textsubscript{2}
very difficult even when taking temperature variations and spectral
doubling into account. We find no significant detections of SO\textsubscript{2}
to greater than 1$\sigma$ confidence. 

\begin{figure}[h]
\includegraphics[width=1\columnwidth]{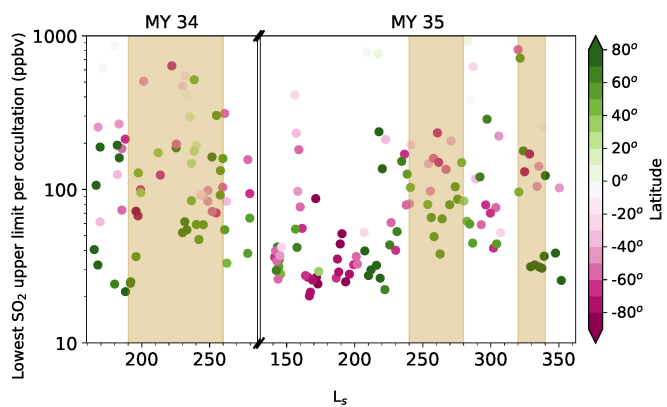}

\caption{Change in the distribution of the lowest SO\protect\textsubscript{2}
upper limit values derived from each ACS occultation sequence as a
function of season, with the colour of each circle representing the
approximate latitude value where the measurement was obtained. Brown
shaded regions indicate major dust storm events \citep{montabone2020,olsen2021hcl}.}
\label{so2dist}
\end{figure}

In Fig. \ref{so2dist} we plot the lowest upper limits retrieved per
occultation as a function of altitude and latitude. For reference,
we mark periods of increased dust storm activity during perihelion,
which usually consists of a large dust storm event that affects the general
latitude range between 60$\text{\textdegree}$ S and 40$\text{\textdegree}$ N and
peaking around $L_{s}=220\text{\textdegree}-240\text{\textdegree}$ , followed by a dip
in dust activity just after solstice and a smaller regional dust storm
event around $L_{s}=320\text{\textdegree}$ that mostly affects southern mid-latitudes
\citep{wangrichardson2015,montabone2015}. Dust activity was particularly
intense in MY 34 compared with MY 35 \citep{montabone2020,olsen2021hcl},
and although the poles remained relatively clear of dust even during
the MY 34 global dust storm event we were only able to get upper limits
down to around 50 ppbv in northern polar regions. By contrast, outside
the perihelion dust storm events, the atmosphere could be probed deep
enough to attain sufficient sensitivity to regularly attain 1$\sigma$
upper limits of 20~ppbv down at around 10 km above the areoid, close
to both the northern and southern polar regions. 

\begin{figure*}[t]
\includegraphics[width=1\textwidth]{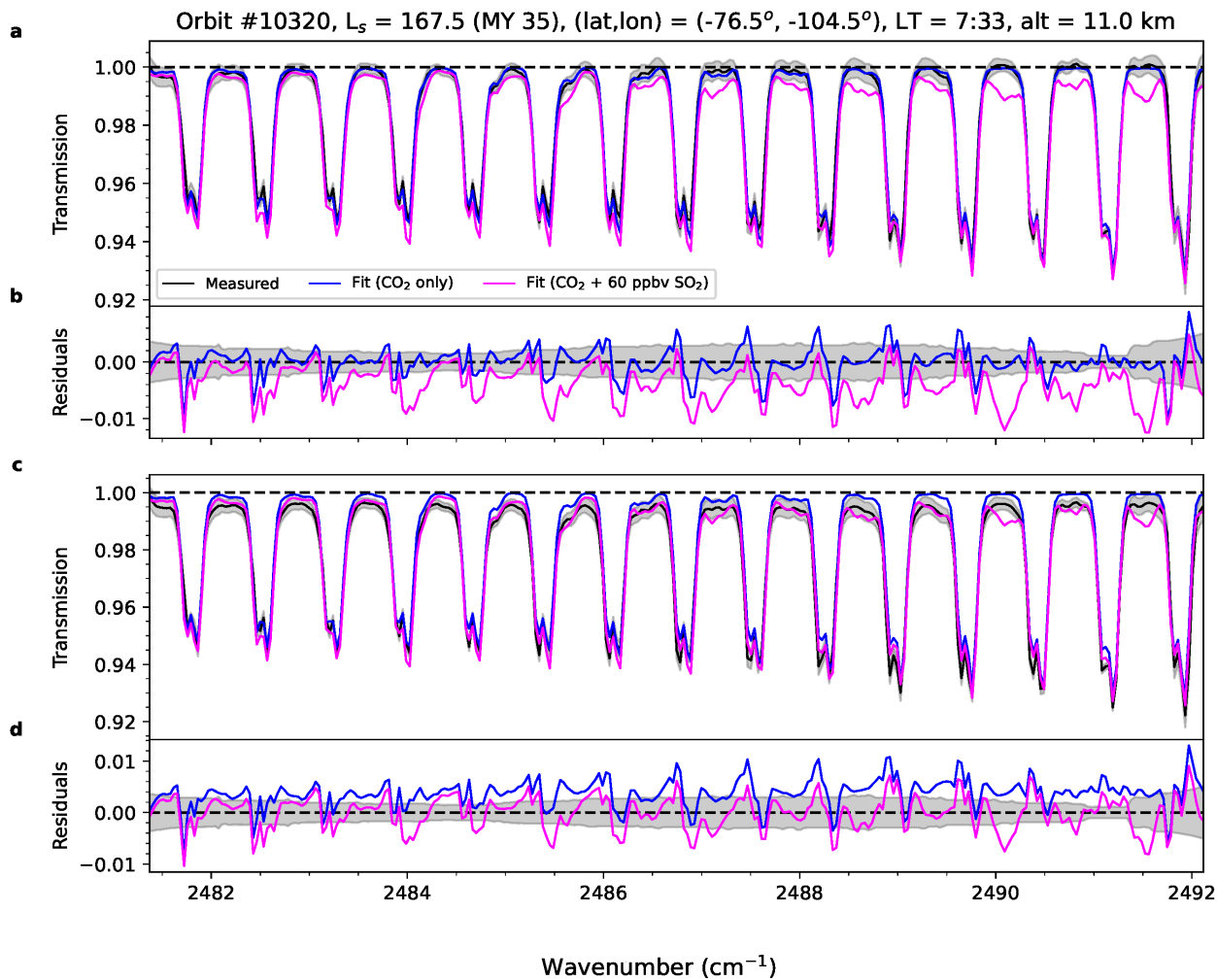}

\caption{Example fit to a position 9 spectrum for which the lowest upper limits
were retrieved from the entire position 9 dataset. In blue is shown
the fit to the measured spectrum taking only CO\protect\textsubscript{2}
absorption and instrumental parameters into account, with the synthetic
spectrum showing the additional contribution of 60 ppbv of SO\protect\textsubscript{2},
corresponding to the perceived 3$\sigma$ level, superimposed in magenta.
Estimated spectral uncertainty is shaded in grey. All spectra are
normalised to the retrieved transmission baseline at unity for clarity.
\emph{Panels a-b: }Spectral fit when the transmission baseline is retrieved
only for the initial fitting of the CO\protect\textsubscript{2} lines
with zero SO\protect\textsubscript{2} abundance. \emph{Panels c-d:
}Spectral fit when the transmission baseline is re-retrieved following
the addition of 60~ppbv of SO\protect\textsubscript{2} into the forward
model.}

\label{so2fits}
\end{figure*}

The spectrum where we were able to find the best upper limit value
of 20 ppbv is shown in Fig. \ref{so2fits}, where we compare the fit
using a fixed abundance of SO\textsubscript{2} in the forward model
that is equal to 3 times the derived 1 sigma upper limit, to the corresponding
fit where no SO\textsubscript{2} is taken into account in the forward
model. The change in the aforementioned doubling effect over the
breadth of the diffraction order is clearly seen in Fig. \ref{so2fits},
with each CO\textsubscript{2} absorption line exhibiting two local
absorption minima where the right minimum progressively dominates
more over the left minimum as one moves towards the right edge of
the detector. The uncertainty induced by this doubling effect on the
transmission baseline ensures that the transmission baseline is difficult
to derive \emph{a priori} without knowledge of SO\textsubscript{2}
abundance, and this is shown clearly in Panel a of Fig. \ref{so2fits}
where fixing the baseline ensures that given abundances of SO\textsubscript{2}
result in SO\textsubscript{2} absorption lines that penetrate further
out of the noise level than if the baseline is allowed to vary as
in Panel c, and hence results in deceptively low SO\textsubscript{2}
upper limits. A better-constrained transmission baseline would result
in upper limits that are approximately halved, which is also reflected
in quoted theoretical upper limits in the literature of 7~ppbv assuming
low dust conditions \citep{korablev2018}. These are still far higher
than the aforementioned upper limits of 0.3 - 2 ppbv retrieved from
stronger absorption bands in the thermal infrared and the sub-millimetre.
Since the uncertainties in the retrievals of SO\textsubscript{2}
are so heavily dominated by sources of systematic error, we cannot
analyse multiple rows of the detector array in order to reduce these
values further. SO\textsubscript{2} is predicted to be well mixed
in the atmosphere below 30 - 50~km altitude after approximately six
months, depending on the volume of outgassing \citep{krasnopolsky1993,krasnopolsky1995,wong2003},
well below the predicted photochemical lifetime of 2 years \citep{nair1994,krasnopolsky1995}.
We therefore find that we can still probe to deep enough altitudes
for any large surface eruption of SO\textsubscript{2} to be monitored
from ACS MIR data within months of its occurrence and be able to track
its origin.

\subsection{H\protect\textsubscript{2}S}

Despite lower predicted outgassing rates of H\textsubscript{2}S compared
with SO\textsubscript{2}, and hence a lower likelihood of detectability,
the concurrent detection of H\textsubscript{2}S is important for
three main reasons. Firstly it is a superior tracer of the location
of outgassing from the surface due to its shorter photochemical lifetime,
with quoted values ranging from of the order of a week \citep{wong2003,wong2005}
to 3 months \citep{summers2002}, well below the timescales for global
mixing in the atmosphere of Mars. Secondly, the H\textsubscript{2}S/SO\textsubscript{2}
ratio provides additional information on the temperature and water
content of an outgassing event, as it is governed by a redox reaction
in which a lower temperature and higher water content favours the
production of H\textsubscript{2}S from SO\textsubscript{2} (\citealt{oppenheimer2011};
and references therein). Finally, H\textsubscript{2}S is, in itself,
a gas that is produced biotically on Earth through the metabolism
of sulphate ions in acidic environments \citep{bertaux2007}, which
could hypothetically be produced by micro-organisms living in sulphate
deposits on Mars.

Even the strongest H\textsubscript{2}S lines in the ACS MIR wavenumber
range are relatively weak, with maximum HITRAN 2016 line strengths
of $\sim$1.6 x 10\textsuperscript{-21}cm\textsuperscript{-1}/(molecule~cm\textsuperscript{-2})
at 296~K, compared with equivalent lines found in the thermal infrared
at 1293 cm\textsuperscript{-1}($\sim$3.3 x 10\textsuperscript{-20}~cm\textsuperscript{-1}/(molecule~cm\textsuperscript{-2});
\citealp{maguire1977,encrenaz2004}) and especially with lines in
the sub-millimetre  ($\sim$4.6x 10\textsuperscript{-20}~cm\textsuperscript{2}~GHz;
\citealt{pickett1998,encrenaz1991,khayat2015})%
\begin{comment}
conversion factor from nm2 MHz is stated in Pickett paper, need to
go back onto database once it stops being down
\end{comment}
. In addition, the spectral region in which most H\textsubscript{2}S
absorption lines are found is dominated by the absorption of several
different isotopologues of CO\textsubscript{2} and H\textsubscript{2}O
\citep{alday2019}, with especially strong absorption by H\textsubscript{2}\textsuperscript{16}O.
We focus here on one particular spectral region between 3827 - 3833~cm\textsuperscript{-1}
around some of the strongest H\textsubscript{2}S bands in the ACS
MIR wavenumber range, located in diffraction order 228 of grating
position 5, as well as an additional spectral window around a single
strong line at 3839.2~cm\textsuperscript{-1} located near the edge
of the diffraction order where the S/N is lowest. Temporal coverage
for this spectral range is restricted to a relatively small number
of observations during the middle (L\textsubscript{s} = 164 - 218$\text{\textdegree}$)
and end (L\textsubscript{s} = 315 - 354$\text{\textdegree}$) of MY 34, the
former providing reasonable spatial and temporal overlap with concurrent
observations of SO\textsubscript{2}.

\begin{figure}[H]
\includegraphics[bb=0bp 0bp 461bp 346bp,width=1\columnwidth]{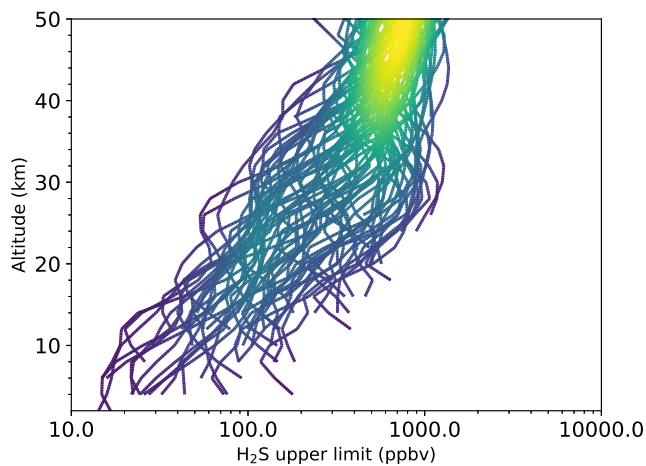}\caption{Retrieved 1$\sigma$ upper limits of H\protect\textsubscript{2}S
from the position 5 dataset as a function of altitude. Colours indicate
the density of retrieved values at each altitude.}
\label{h2sdist}
\end{figure}

\begin{figure*}
\includegraphics[width=0.92\textwidth]{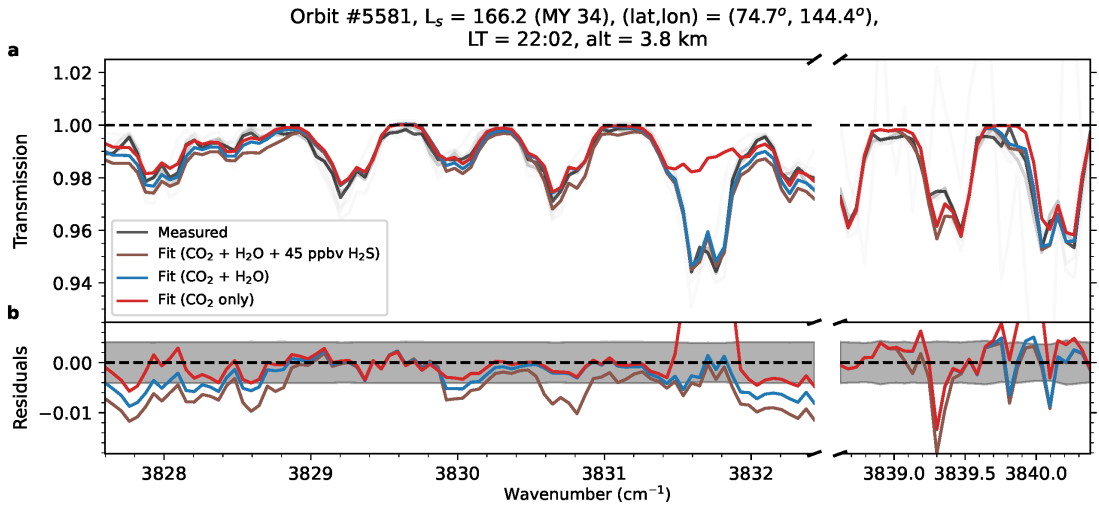}

\caption{Upper limit of H\protect\textsubscript{2}S from position 5 spectra. \emph{ Panel a: }Fit to the spectrum with the lowest retrieved H\protect\textsubscript{2}S
upper limit in the position 5 dataset, with a 3$\sigma$ value equivalent
to 45 ppbv. In grey are the observed spectra at each row of the detector
array normalised to the retrieved unity baseline (black, dashed):
the darker the grey colour, the greater the weighting used for the
calculation of $\mu^{*}$ and $\varsigma^{*}$. The red and blue lines
respectively show the contributions of each of the interfering gas
species (in this case, CO\protect\textsubscript{2} and H\protect\textsubscript{2}O)
to the fit to the row of the detector array with the greatest S/N,
while the brown line shows the synthetic spectrum following the addition
of 3$\sigma$ of H\protect\textsubscript{2}S. \emph{Panel b: }Residuals
of each of the gas contributions compared with the observed spectrum
with the greatest S/N, where the shaded grey region represents the
estimated noise level.}

\label{h2sfits}
\end{figure*}

In Fig. \ref{h2sdist} we show the retrieved H\textsubscript{2}S
upper limit profiles from all 86 processed position 5 observations.
Although the sample size is small, we clearly see that upper limit
values down to at least 30~ppbv can be achieved relatively consistently
around 20~km altitude. In the best case we were able to achieve an
upper limit down to 15~ppbv at 4~km altitude, which we show in Fig.
\ref{h2sfits}. From our spectral fits it is clear that such an upper
limit is mostly imposed by the 3830.7~cm\textsuperscript{-1} and
3831~cm\textsuperscript{-1} H\textsubscript{2}S absorption lines,
the latter of which is the only one fully resolvable from neighbouring
CO\textsubscript{2} and H\textsubscript{2}O lines and less affected
by the uncertainty on the ACS MIR doubling instrument function. While
this 15~ppbv value is in line with expected values from \citet{korablev2018},
who quote predicted upper limits of 17~ppbv for ACS MIR in low dust
conditions, it is still far higher than the 1.5~ppbv obtained by
\citet{khayat2015} in the sub-millimetre. Experimental evidence from the
Curiosity rover found that heating samples of regolith to very high
temperatures would result in the emission of both SO\textsubscript{2}
and H\textsubscript{2}S at an H\textsubscript{2}S/SO\textsubscript{2}
ratio of approximately 1 in 200 \citep{leshin2013}. This indicates
that, given our constraints on SO\textsubscript{2} abundance, passive
outgassing is unlikely to lead to volume mixing ratios of H\textsubscript{2}S
in the troposphere above around 0.4~ppbv, far below the detection
capabilities of ACS MIR.

\subsection{OCS}

\begin{figure}[h]
\includegraphics[width=1\columnwidth]{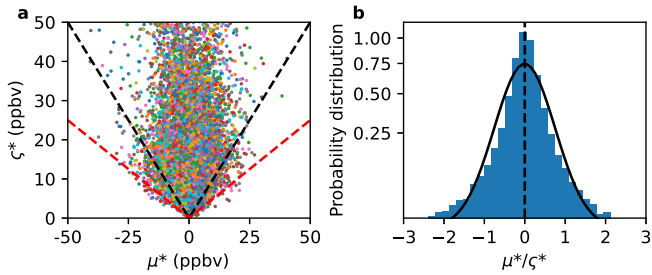}

\caption{Statistical distribution of the retrieved values of $\mu^{*}$ and
$\varsigma^{*}$ for OCS from all position 11 spectra in the analysed
dataset. \emph{Panel a:} Scatter plot of $\mu^{*}$ against $\varsigma^{*}$,
with the dashed black line showing the 1$\sigma$ detection threshold
and the dashed red  line showing the 2$\sigma$ detection threshold.
\emph{Panel b: }Normalised probability distribution of derived $\sigma$
detection values, with the solid black line showing the Gaussian best
fit to the probability distribution. While we find little skew in
our upper limit retrievals, there is some kurtosis due to occasional
overfitting of fixed pattern noise.}

\label{ocsnormaldist}
\end{figure}

\noindent 
Although predicted outgassing of OCS is much lower than for SO\textsubscript{2}
and H\textsubscript{2}S, with the OCS/SO\textsubscript{2} ratio
from terrestrial volcanic emission observed to be of the order of
10\textsuperscript{-4} - 10\textsuperscript{-2} (e.g. \citealp{sawyer2008,oppenheimerkyle2008}),
it could nonetheless be produced indirectly from volcanic SO\textsubscript{2}
in the high atmosphere through reaction with carbon monoxide \citep{hongfegley1997}.
Older systematic searches for OCS from Mariner 9 data \citep{maguire1977}
and ground-based millimetre observations \citep{encrenaz1991} established
upper limits in the Martian atmosphere of 70~ppbv. More recently,
\citet{khayat2017} stated upper limits of 1.1~ppbv from ground-based
observations of a mid-infrared band centred at 2925~cm\textsuperscript{-1},
which was partially obscured by telluric absorption. This band is
also the strongest of two OCS absorption bands that is covered by
ACS MIR, specifically by grating position 11 (the other, centred around
3460 - 3500~cm\textsuperscript{-1} and covered by grating position
3, is weaker and contaminated by strong CO\textsubscript{2} absorption).
It also has by far the best spatial and temporal coverage out of all
the three grating positions shown here, providing almost continuous
coverage from the start of the ACS science phase in April 2018
(MY 34, L\textsubscript{s} = 163\textdegree) to January 2021 (MY 35, L\textsubscript{s}
= 355\textdegree). Unlike with SO\textsubscript{2} and H\textsubscript{2}S
there is a relative lack of other gases that absorb in this region
and would further complicate the retrieval of OCS, with the exception
of HCl, which is present only during certain seasons and which is easily
isolated from the OCS absorption lines. On the flip side, the lack
of gases that absorb in this region can also make it difficult to
fit the instrument line shape associated with spectral doubling, especially
for occultations where there is an absence of HCl. This can also occasionally
result in overfitting of OCS to local noise features present in the
spectrum, resulting in spurious detections usually of the order of
2$\sigma$. We show this in figure \ref{ocsnormaldist}, where the
retrieved ratios of $\mu^{*}/\varsigma^{*}$ for all position 11 measurement
sequences and altitudes should statistically approximate a Gaussian
distribution centred around $\mu^{*}/\varsigma^{*}\approx0$ and with
a standard deviation equivalent to $\mu^{*}/\varsigma^{*}$, but we
instead find some positive kurtosis due to the presence of false 2$\sigma$
detections. Nonetheless, the effect of overfitting noise is somewhat
mitigated through averaging over adjacent rows on the detector array,
and we find no evidence of OCS in our observations above a 3$\sigma$
confidence level.

\begin{figure}[h]
\includegraphics[width=1\columnwidth]{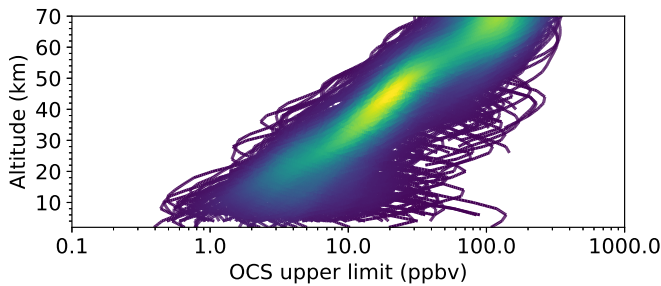}\caption{Retrieved 1$\sigma$ upper limit values of OCS from all position 11
measurement sequences analysed in the dataset. Colours indicate the
density of upper limit values as in Figs. \ref{so2scatter} and \ref{h2sdist}. }

\label{ocsscatter}
\end{figure}

\begin{figure}[h]
\includegraphics[width=1\columnwidth]{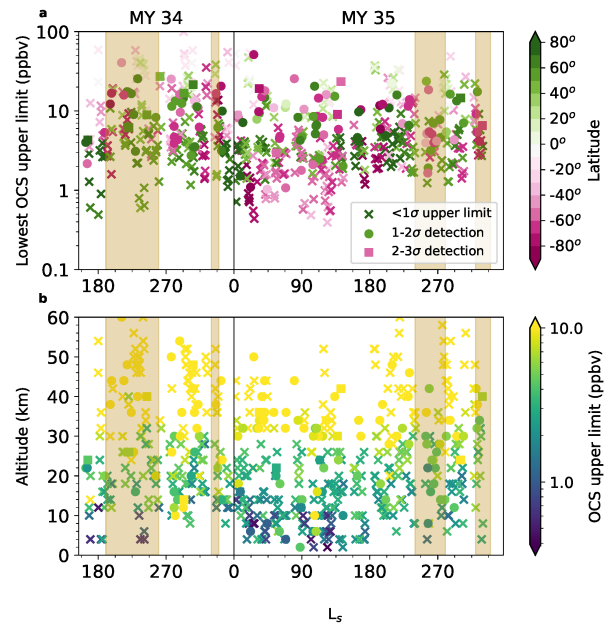}

\caption{Seasonal distribution of lowest OCS upper limit values from each position
11 measurement sequence, as a function of latitude (\emph{Panel a})
and altitude (\emph{Panel b}). Crosses denote upper limits of below
1$\sigma$ significance, circles denote detections of between
1$\sigma$ and 2$\sigma$, and squares denote detections of between 2$\sigma$ and 3$\sigma.$
Brown shaded regions indicate major dust storm events, as in Fig. \ref{so2dist}.
No positive detections of OCS above 3$\sigma$ were found in the data.}

\label{ocsdists}
\end{figure}

\begin{figure*}
\includegraphics[width=1\textwidth]{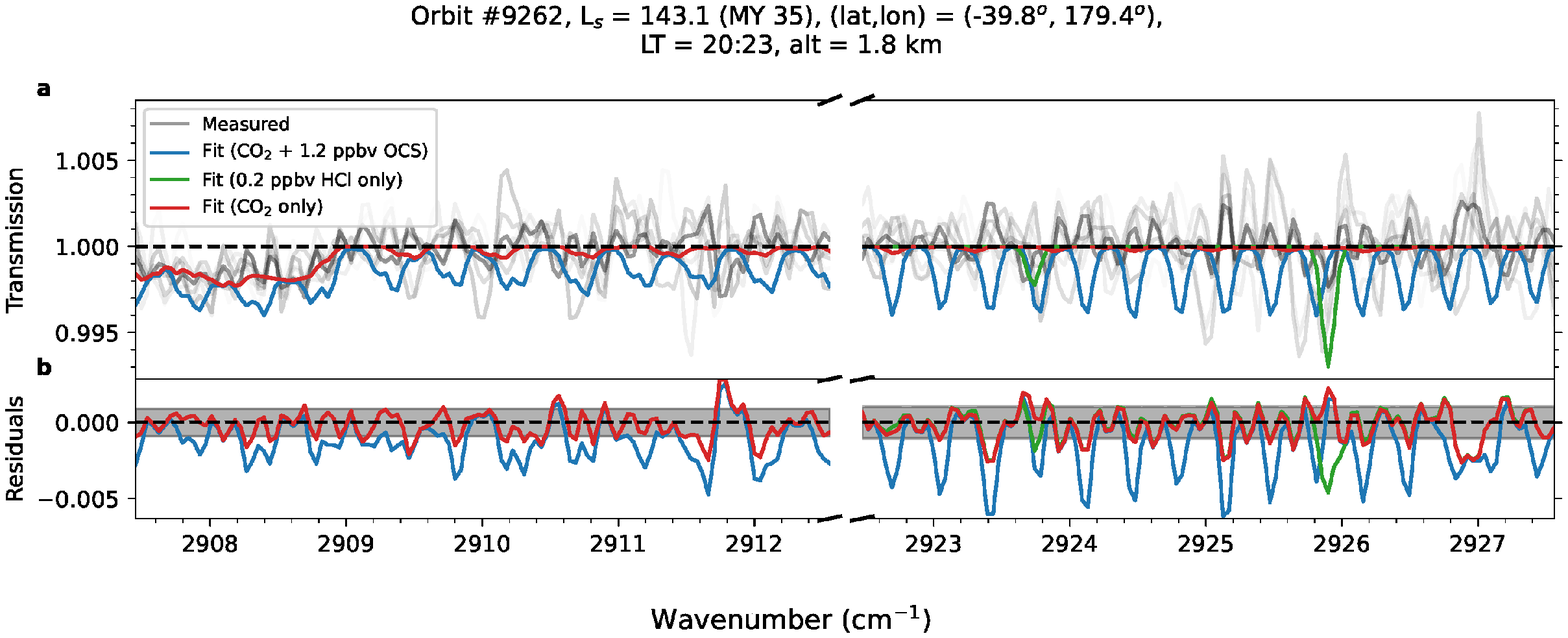}

\caption{Upper limit of OCS from position 11 spectra. \emph{Panel a: }Fit to the spectrum with the lowest retrieved OCS
upper limit in the position 11 dataset, with a 3$\sigma$ value equivalent
to 1.2 ppbv. As in Fig. \ref{h2sfits}, the grey lines represent each
of the spectra on the detector array used to estimate the OCS detection
limit, where the darker the grey colour, the higher the S/N. On the
left we show the fit to the spectrum in diffraction order 173, and
to the right we show the simultaneous fit to order 174. For reference,
we also plot the approximate 3$\sigma$ upper limit of HCl in green,
equivalent to 0.2 ppbv, as well as the contribution of CO\protect\textsubscript{2}
to the fit. \emph{Panel b: }Residuals of the spectral fit.}

\label{ocsfits}
\end{figure*}

As with SO\textsubscript{2}, the greatest density of retrieved upper
limits can be found around 20 km of altitude as shown in Fig. \ref{ocsscatter},
where values of 2-3~ppbv can regularly be sought even in regions
of high dust concentration. Unlike with SO\textsubscript{2}, however,
observations at lower altitudes are much less limited by the presence
of systematic uncertainties or aerosols, which allows for smaller
upper limit values to be sought much closer to the surface. In Fig.
\ref{ocsdists} we show that the lowest upper limits can be obtained
at the winter poles down to 0.4~ppbv, where the atmosphere can be
probed by ACS MIR down to around 2~km above the areoid, and especially
the southern hemisphere during winter solstice where global or regional
dust activity is minimal, for which the best example is shown in Fig.
\ref{ocsfits}. By contrast, the perihelion dust season usually limits
detections to above 1~ppbv, in line with previous upper limit estimates
from \citet{khayat2017} that were obtained through co-adding of both
aphelion and perihelion observations. This is also an improvement
on theoretical upper limit estimates previously predicted for ACS
MIR by \citet{korablev2018}, where the expected performance of the
instrument only predicted values down to 2~ppbv even in clear atmospheric
conditions. Nonetheless, if we were to attempt to search for real
OCS outgassing given an upper limit constraint of 20~ppbv provided
by our SO\textsubscript{2} retrievals, together with the constraints
on the OCS/SO\textsubscript{2} provided by terrestrial volcanoes,
even if we were to assume a very high ratio of OCS/SO\textsubscript{2}=
10\textsuperscript{-2} we would realistically require sensitivity
to OCS below at least 0.2~ppbv. This is also complicated by the fact
that OCS is difficult to form at low temperatures according to the
\citet{hongfegley1997} reaction mechanism and would require very
hot sub-surface temperatures to be present on Mars. \citet{oppenheimer2011}
quote OCS/SO\textsubscript{2} ratios for the Antarctic volcano Erebus
of $7\times10^{-3}$, equivalent to a maximum of 0.15~ppbv of OCS
given detected SO\textsubscript{2} values on Mars. All these criteria
make it very unlikely that ACS MIR would be able to find OCS in the
Martian atmosphere in sufficiently high quantities to be detected.


\section{Discussion and conclusion}

In this analysis we present the results of a systematic study of multiple
solar occultation observations of Mars during Martian years 34 and
35 using ACS MIR,
with the aim of either detecting or establishing upper limits on the
presence of three major signatures of volcanic outgassing. For SO\textsubscript{2},
we constrain gas abundances to below 20~ppbv. Assuming that SO\textsubscript{2}
is well mixed in the atmosphere following an eruption that happened
more than six months prior, and assuming a total mass of the atmosphere
of approximately $2.5\times10^{16}$~kg%
\begin{comment}
better citation than NASA fact sheet?
\end{comment}
, this translates to a total maximum limit of approximately 750~ktons
of SO\textsubscript{2} in the atmosphere, averaging an outgassing
rate of less than 2 ktons a day if we assume that SO\textsubscript{2}
has a lifetime in the Martian atmosphere of approximately 2 years.
For comparison, the most active volcanoes on Earth -- Etna in Sicily
and Kilauea in Hawaii -- have passive outgassing rates of SO\textsubscript{2}
of approximately 5.5~ktons/day and 0.98~ktons/day, respectively (\citealt{oppenheimer2011};
and references therein), while a single eruption of a Volcanic Explosivity
Index (VEI) of 3, equivalent to a small eruption that is expected
to occur approximately once every few months on Earth, can be expected
to emit around 700~ktons of SO\textsubscript{2} into the atmosphere
\citep{graf1997volcanic}. This appears to reinforce the prevailing
view that residual present-day volcanic activity on Mars can only
exist at an extremely low level, if at all. In addition, we derive
H\textsubscript{2}S upper limits down to 16~ppbv and OCS upper
limits down to 0.4~ppbv in the best cases, the latter value being
lower than any values previously published. However, these are unlikely
to be of sufficient sensitivity for ACS MIR to detect passive outgassing
from the surface of Mars. For all three molecules, no positive detections
beyond 3$\sigma$ were found to be present in the ACS MIR data.

The lack of sulphur compounds detected in the Martian atmosphere has
implications for the origin of two other molecules that have been
detected in recent years. Halogen halides (HF, HCl, HBr, and HI) are
known to be emitted by terrestrial volcanoes, and only one, HCl, has
so far been confirmed to exist in the Martian atmosphere. HCl was
seen to peak in abundances of the order of a few ppbv in the second
half of both MY 34 and MY35, between the global and regional dust storm
periods \citep{korablev2020hcl,olsen2021hcl}. The HCl/SO\textsubscript{2}
mass ratio in gases emitted from terrestrial volcanoes is variable
but usually lies between around 0.1 - 0.9 \citep{pylemather2009}. This would be just compatible with the SO\textsubscript{2} upper
limits found in this work, although seasonal trends in HCl concentration
appear not to favour a primarily volcanic origin. With regards to
the origin of methane, the ratio of CH\textsubscript{4}/SO\textsubscript{2}
emitted by terrestrial volcanoes is 1.53 at the most, and usually
much smaller (\citealt{nakagawa2009}; and references therein). The
recent 20~ppbv and 45~ppbv methane spikes claimed by \citet{moores2019}
and \citet{mumma2009} would therefore only be partially justifiable
as being of volcanic origin, assuming they are confirmed to be genuine.

ACS MIR observations of SO\textsubscript{2} and H\textsubscript{2}S
have so far been hampered by a lack of spatial and temporal coverage,
as well as an instrument line shape function that as yet remains incompletely
characterised. Additional measurements in the future could allow us
to probe deeper into the atmosphere and drive down upper limits on SO\textsubscript{2}
and H\textsubscript{2}S even further to definitively constrain the
amount of outgassing from the surface of Mars, especially around regions
near the tropics such as Cerberus Fossae, where signs of intermittent
volcanism are the most promising.
\begin{acknowledgements}
The ACS investigation was developed by the Space Research Institute
(IKI) in Moscow, and the Laboratoire Atmosph\`eres, Milieux, Observations
Spatiales (LATMOS) in Guyancourt, France. The investigation was funded
by Roscosmos and the French National Centre for Space Studies (CNES). This work was
funded by CNES, the Agence Nationale de la Recherche (ANR, PRCI, CE31
AAPG2019, MCUBE project), the Natural Sciences and Engineering Research
Council of Canada (NSERC) (PDF\textendash 516895\textendash 2018),
the UK Space Agency and the UK Science and Technology Facilities Council
(ST/T002069/1, ST/R001502/1, ST/P001572/1). AT, OIK and AAF were funded by the Russian Science Foundation (RSF) 20-42-09035 for part of
their contribution described below. All ACS MIR spectral fitting
was performed by ASB, while ACS NIR
 spectral fitting of pressure-temperature
profiles was performed by AAF and GCM-derived pressure-temperature
profiles were created by FF and EM. The interpretation of all results
in this work was done by ASB, AT, FM and KSO. Pre-processing and calibration
of ACS spectra was performed at IKI by AT and at LATMOS by LB. Spatio-temporal
metadata were produced in LATMOS by GL and in IKI by AP. Input and
aid on spectral fitting were given by JA, LB, FM, KSO and AT. The
ACS instrument was designed, developed, and operated by OIK, FM, AP,
AS and AT.
\end{acknowledgements}

\bibliographystyle{aa}
\bibliography{42390corr_arxiv}

\appendix

\section{Detailed procedure of upper limit derivation for H\protect\textsubscript{2}S
and OCS\label{sec:Detailed-procedure-of}}

We assume that in the case where the trace gas cannot be identified
from the spectra, the volume mixing ratio of the trace gas measured
at each altitude follows a Gaussian distribution centred at a zero
abundance value, with a standard deviation corresponding to the uncertainty
in the retrieved profile due to both spectral noise and sources of
systematic error as per \citet{montmessin2021} and \citet{knutsen2021}.
This means that we want to ensure that there is as little prior constraint
on the vertical profile as possible, and ensure that vertical correlations
are sufficiently small to induce oscillations in the retrieved profiles
that approximately reflect the sigma uncertainty level at each altitude.
The six-step procedure is summarised as follows.\ 

Step 1: We estimated the S/N from random noise by measuring the degree of fluctuation
around local maxima present in each transmission spectrum (method
3 in \citealp{braude2021soar}).

Step 2: We performed a preliminary retrieval of the trace species abundance in
question, together with any other instrumental and atmospheric parameters,
allowing as much variability as possible. %
\begin{comment}
starting from a prior value of 0 at all altitudes, and make the prior
errors as large as possible so that the retrieval is constrained more
by the noise level than the prior.
\end{comment}

Step 3: In order to estimate the contribution of spectral uncertainty from
sources of systematic error that are difficult to quantify (such as
forward modelling error or local calibration artefacts), we divided the
S/N estimated in step 1 at each tangent height by the square root
of the reduced $\chi^{2}$ value obtained at each tangent height,
analogously to \citet{montmessin2021}. 

Step 4: In order to further reduce biases from the prior and errors in convergence
due to ill conditioning, we performed a final retrieval with the prior
trace gas profile being the retrieved profile from step 3 multiplied
by -1, and with the prior errors on the trace gas profile reduced
by an order of magnitude. This also allows a detection to be more
easily distinguished from a non-detection. %
\begin{comment}
This is done in order to minimise the likelihood that the profile
would repeatedly converge to a given value simply due to biases in
the retrieval itself, as opposed to genuine spectral features. In
addition, it forces the distribution of retrieved values at each altitude
to more closely approximate a Gaussian distribution where the number
of retrievals per dataset is too small for the Central Limit Theorem
to apply. Hence, in the case of a negative detection, the two retrieved
profiles together would be approximately symmetrical around 0, while
in the case of a positive detection, the two retrieved profiles would
converge close to the same value.
\end{comment}
\begin{comment}
might need to show this pictorally
\end{comment}

Step 5: We repeated steps 1 - 4 for approximately seven rows on the detector array
for a given diffraction order, with the rows chosen to sample the
change in the S/N over the detector array. %
\begin{comment}
This reduces the likelihood of mistaking a local artefact or noise
feature for a genuine detection.
\end{comment}

Step 6: The detection limit at each altitude, $j$, was then the weighted mean,
$\mu_{j}^{*}$, plus the weighted standard deviation, $\varsigma_{j}^{*}$,
of the trace gas abundance values, $x_{ijk}$, retrieved from each row,
\emph{i}, of the detector array\emph{, }with $k=1$ corresponding to
the value retrieved from step 3 and $k=2$ corresponding to that retrieved
from step 4\emph{ }(with \emph{a posteriori} variance $\Delta_{ijk}$
on the value of $x_{ijk}$ determined through propagation of errors
in each retrieval):

\begin{center}
\begin{equation}
\mu_{j}^{*}=\frac{\sum_{k=1}^{2}\sum_{i=1}^{N}\frac{x_{ijk}}{\Delta_{ijk}}}{\sum_{k=1}^{2}\sum_{i=1}^{N}\frac{1}{\Delta_{ijk}}}
\end{equation}
\par\end{center}

\begin{center}
\begin{equation}
\varsigma_{j}^{*}=\sqrt{\frac{\sum_{k=1}^{2}\sum_{i=1}^{N}\frac{(x_{ijk}-\mu_{j}^{*})^{2}}{\Delta_{ijk}}}{\left(\sum_{k=1}^{2}\sum_{i=1}^{N}\frac{1}{\Delta_{ijk}}\right)-\frac{\left(\sum_{k=1}^{2}\sum_{i=1}^{N}\frac{1}{\Delta_{ijk}}^{2}\right)}{\left(\sum_{k=1}^{2}\sum_{i=1}^{N}\frac{1}{\Delta_{ijk}}\right)}}}
,\end{equation}
\par\end{center}

\noindent where \emph{N }is the number of detector rows sampled in the diffraction
order. The weighting factor $\frac{1}{\Delta_{ijk}}$implicitly takes
into account the fact that the noise profile, S/N and the quality
of the spectral fit vary from row to row, and acts to maximise the
contribution of rows where the fit is optimal and noise is reduced.
Hence the parameter $\varsigma_{j}^{*}$ can be thought of as the
total uncertainty on the retrieved gas profile due to random noise,
while the parameter $\mu_{j}^{*}$ represents total systematic offsets
in the retrieved gas profile that are either due to genuine absorption
lines or due to systematic uncertainties such as calibration artefacts
or forward modelling error.
\end{document}